\documentclass[%
reprint,
aip,
showkeys,
amsmath,amssymb,
longbibliography,
]{revtex4-1}

\usepackage{graphicx}
\usepackage[caption=false]{subfig} 
\usepackage{pifont}

\usepackage{dcolumn}
\usepackage{bm}

\usepackage[mathscr]{euscript}  
\usepackage{xcolor}  

\usepackage{tcolorbox}
\usepackage{comment}
\usepackage{float}

\makeatletter
\newsavebox{\@brx}
\newcommand{\llangle}[1][]{\savebox{\@brx}{\(\m@th{#1\langle}\)}%
	\mathopen{\copy\@brx\kern-0.5\wd\@brx\usebox{\@brx}}}
\newcommand{\rrangle}[1][]{\savebox{\@brx}{\(\m@th{#1\rangle}\)}%
	\mathclose{\copy\@brx\kern-0.5\wd\@brx\usebox{\@brx}}}
\makeatother

\makeatletter
\newcommand{\vast}{\bBigg@{3.4}}
\makeatother

\usepackage{tikz}

\makeatletter
\newcommand\emailx[1]{%
	\move@AF%
	\def\@affil{{\normalfont\,#1\strut}{}}%
}%

\begin{document}

	\preprint{ApS/123-QED}

	\title{Dynamical density functional theory for dense odd-diffusive fluids}

	\author{Iman Abdoli}
	\email{iman.abdoli@hhu.de}
	\affiliation{Institut für Theoretische Physik II - Weiche Materie, Heinrich-Heine-Universität Düsseldorf, Universitätsstraße 1, D-40225 Düsseldorf, Germany}
	
	\author{Ren\'e Wittmann}
	\affiliation{ Max Rubner-Institut (MRI), Institut f\"ur Sicherheit und Qualit\"at bei Fleisch, D-95326 Kulmbach, Germany}
	\affiliation{Institut für Theoretische Physik II - Weiche Materie, Heinrich-Heine-Universität Düsseldorf, Universitätsstraße 1, D-40225 Düsseldorf, Germany}

	\author{Hartmut L\"{o}wen}
	\email{Hartmut.Loewen@uni-duesseldorf.de}
	\affiliation{Institut für Theoretische Physik II - Weiche Materie, Heinrich-Heine-Universität Düsseldorf, Universitätsstraße 1, D-40225 Düsseldorf, Germany}

\begin{abstract}
Odd diffusion breaks time-reversal symmetry in overdamped systems through transverse probability currents while preserving equilibrium steady states. In this work, we develop a dynamical density functional theory (DDFT) for dense interacting odd-diffusive fluids and apply it to ultrasoft particles in two dimensions.
In bulk, odd diffusion qualitatively reshapes collective relaxation by
generating transient circulating current patterns
which do not exist in normal fluids.
Under harmonic ring confinement, the circulation of probability current induces an angular redistribution of density along the ring during relaxation. This unique footprint of odd diffusion opens up a shorter pathway to equilibrium.
Repulsive interactions significantly enhance these effects. 
Excellent agreement with Brownian dynamics simulations confirms that our odd-DDFT framework quantitatively captures all essential nonequilibrium aspects of the nontrivial odd transport and collective redistribution for dense fluids in both bulk and confined geometries.

\end{abstract}
	
	\maketitle

	\section{Introduction}
	
Odd diffusivity generalizes Fickian transport by allowing the diffusivity tensor to acquire an antisymmetric component, in which oddness is encoded by a dimensionless parameter $\kappa$, in two dimensions, such that density gradients generate transverse, Hall-like probability currents in addition to the usual down-gradient fluxes~\cite{PhysRevLett.127.178001}. Microscopically, odd diffusivity arises from the simultaneous breaking of time-reversal and parity symmetries at the level of fluctuations~\cite{hargus2020time}, for instance in chiral random motion~\cite{kummel2013circular, van2008dynamics, vuijk2022active, muzzeddu2022active}, charged Brownian particles in external magnetic fields~\cite{czopnik2001brownian, chun2018emergence, vuijk2019anomalous, PhysRevE.101.012120, abdoli2020correlations, shinde2022strongly, PhysRevE.111.025412}, diffusing skyrmions~\cite{schutte2014inertia, troncoso2014brownian, wiesendanger2016nanoscale, fert2017magnetic, buttner2018theory, weissenhofer2021skyrmion}, or passive objects immersed in chiral active baths~\cite{hargus2025odd}. A unifying theoretical framework has shown that odd diffusivity can be derived from Green--Kubo relations involving velocity cross-correlations and that it is fully compatible with equilibrium steady states: the antisymmetric contribution generates divergence-free currents in bulk and therefore does not modify stationary density distributions
\cite{PhysRevLett.127.178001, vega2022diffusive}.
In addition to odd-diffusive transport, antisymmetric response coefficients arise in systems with odd viscosity~\cite{banerjee2017odd, han2021fluctuating, markovich2021odd, zhao2022odd, lier2023lift}, odd elasticity~\cite{scheibner2020odd, braverman2021topological}, and odd viscoelasticity~\cite{banerjee2021active, lier2022passive}. In experimentally relevant realizations, oddness manifests itself through a rich variety of collective dynamical regimes~\cite{tan2022odd, soni2019odd, bililign2022motile}.

When interactions are introduced, odd diffusion becomes a genuinely collective phenomenon that qualitatively alters how particles rearrange during collisions and crowding. In dilute and moderately dense regimes, purely repulsive interactions combined with odd transport can enhance self-diffusion rather than suppress it, in stark contrast to normal diffusive fluids
\cite{PhysRevLett.129.090601,kalz2024oscillatory,kalz2025reversal}.
Exact two-body and tracer-level analyses reveal that antisymmetric diffusion tensors render the Smoluchowski operator non-Hermitian, leading to oscillatory force correlations, nonmonotonic relaxation, and interaction-induced reversals of longitudinal and transverse responses, even in overdamped equilibrium systems. These effects demonstrate that odd diffusion fundamentally modifies interaction-induced transport already at low densities, while remaining consistent with detailed balance at the level of stationary distributions.
	
Experiments on two-dimensional chiral fluids composed of rotating units have directly measured diffusion tensors with sizable antisymmetric components—often comparable in magnitude to the symmetric diffusivity—and have shown that the sign and strength of odd diffusion are controlled by global flow vorticity rather than microscopic particle activity alone
\cite{vega2022diffusive}.
Complementary theoretical and numerical studies demonstrate that odd transport can reshape collective relaxation, promote circulating probability currents, and generate unconventional dynamical states, including vortex-like motion and pattern formation, even in systems with purely repulsive interactions
\cite{muzzeddu2025self,langer2024dance,mecke2024emergent}.
When promoted to effective odd interactions or combined with inertia, odd transport can further induce spontaneous circulation, bubble-like inhomogeneous phases, and circulating edge currents stabilized by transverse forces
\cite{caprini2025spontaneous,caprini2025odd,caprini2025bubble}.

Despite the significant progress reviewed above, most existing studies of odd diffusion and odd transport focus either on single-particle dynamics, tracer motion, or low-order correlation functions, while the investigation of dense systems usually relies on
kinetic descriptions, that do not resolve the full spatiotemporal evolution of interacting density fields, or particle-based simulations. While these approaches have firmly established that antisymmetric transport coefficients reshape relaxation pathways and probability currents without modifying equilibrium steady states, they do not provide a closed, field-level description of collective density dynamics in interacting systems. In particular, the interplay between odd diffusion, interparticle interactions, and spatial inhomogeneity—such as confinement, external fields, or geometry-induced gradients—remains difficult to access systematically within tracer-based or few-body frameworks. This gap naturally calls for a sophisticated theoretical approach that 
retains a clear microscopic foundation and is capable of characterizing both structural and flow properties 
during nonequilibrium relaxation, while being tractable for practical calculations.

A natural and widely used framework  for the collective dynamics of interacting Brownian particles 
is \emph{dynamical density functional theory} (DDFT)~\cite{te2020classical}, which preserves the microscopic realism of the many-body Smoluchowski equation while operating at the level of one-body fields. 
Originally formulated by Marconi and Tarazona~\cite{marconi_tarazona_1999,marconi_tarazona_2000}, DDFT has since been systematically developed to describe various interacting colloidal systems under external fields, time-dependent driving, and generalized transport mechanisms~\cite{archer_evans_2004, archer2009dynamical,rex2007dynamical,rex2008dynamical,rex2009dynamical,loewen_hydrodynamics_review,  brader2011density, scacchi2016driven, wittmann2021order}. In general, the theory provides a deterministic evolution equation for the one-body density after closing the Smoluchowski hierarchy. Its central approximation is the adiabatic assumption, whereby nonequilibrium correlation functions are replaced by those of an auxiliary equilibrium system with the same instantaneous density profile~\cite{espanol2009derivation}. This closure allows equilibrium Helmholtz free-energy functionals to be employed as microscopic input for nonequilibrium dynamics. For systems of ultrasoft or weakly correlated particles~\cite{likos_review}, the excess free energy is accurately captured by a simple mean-field form involving only the pair potential~\cite{evans1979nature,dzubiella2003mean, archer2017standard}, yielding particularly transparent and versatile DDFT approaches that have been successfully applied to confinement, crystallization, driven relaxation, and complex transport phenomena~\cite{van2008colloidal,van2008crystallization,goh2023density, scacchi2018flow}.

	\begin{figure}
		\centering
		\includegraphics[width=\linewidth]{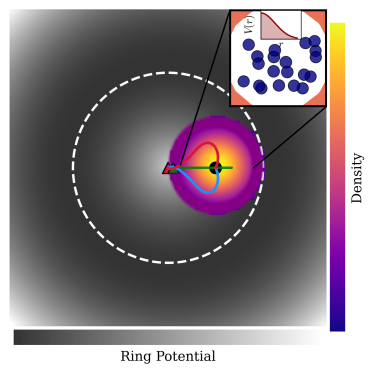}
		\caption{\textbf{Schematic illustration of the confined odd-fluid in a potential ring}. A two-dimensional assembly of ultrasoft Brownian particles is confined by a radially symmetric harmonic ring potential $V_{\mathrm{ext}}(r)$ (gray background), whose minimum defines the preferred radius $R_0$ (white dashed circle). The system is initialized with an off-center Gaussian density distribution (localized blob), which relaxes toward the ring. Superimposed is the trajectory of the center of mass of the density distribution during relaxation for three representative values of the odd-diffusion parameter: $\kappa>0$ (red), $\kappa<0$ (blue), and $\kappa=0$ (green). Odd diffusion induces a transverse drift of the density center of mass, whose direction reverses upon changing the sign of $\kappa$, while purely relaxational motion is recovered for $\kappa=0$. \textit{Inset:} zoom-in of the initial density distribution, illustrating interacting particles governed by an ultrasoft Gaussian pair potential $V(r)=\varepsilon \exp(-r^2/\sigma^2)$, where $\varepsilon$ controls the interaction strength and $\sigma$ the interaction range.}		
		\label{figure1}
		
	\end{figure}

	\begin{figure*}[t]
		\centering
		\includegraphics[width=\textwidth]{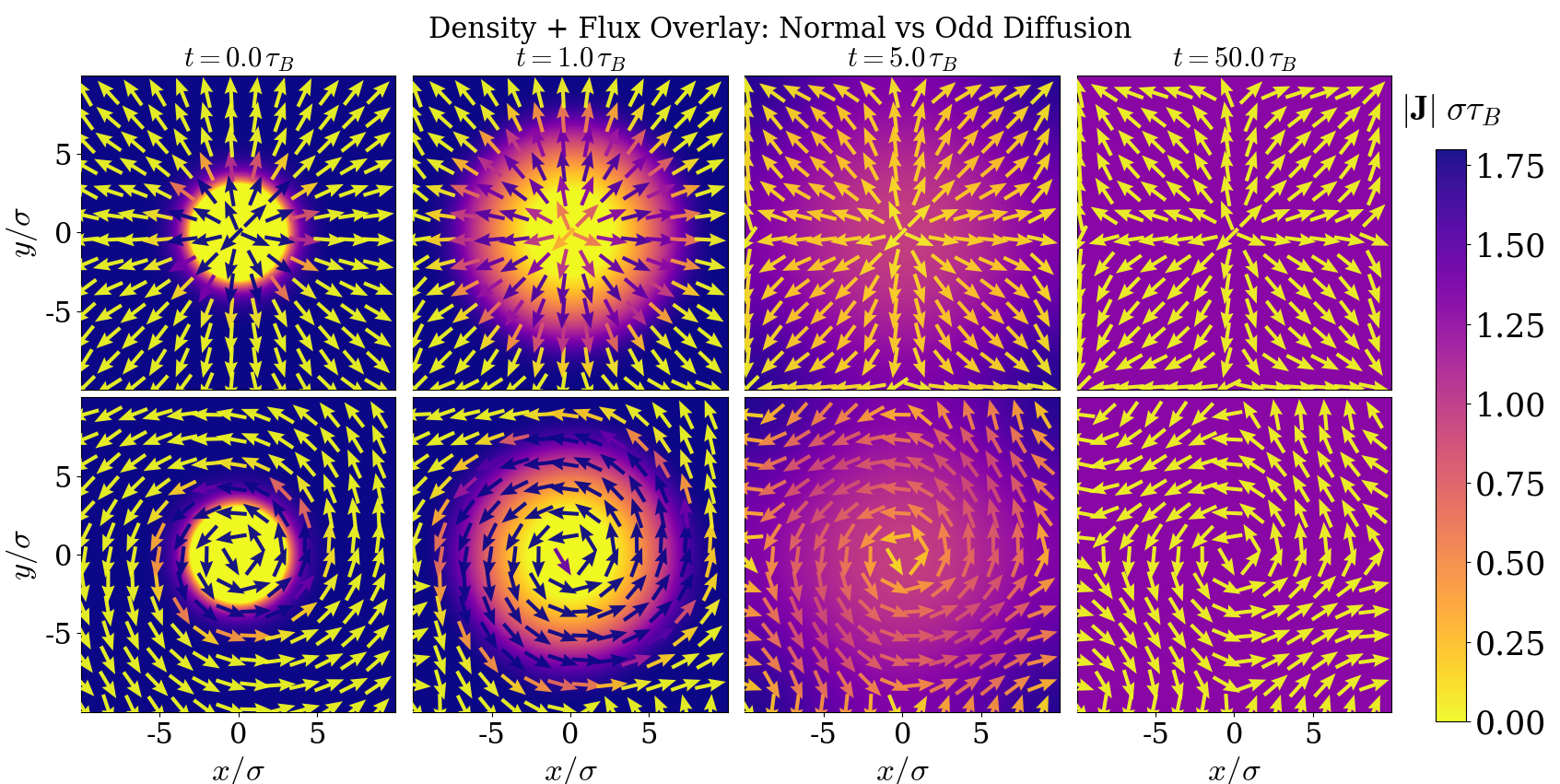}
		\caption{\textbf{Density and flux evolution in a nonconfined interacting fluid for normal and odd diffusion.}
			The system consists of ultrasoft particles interacting via a Gaussian pair potential and is initialized with a centered Gaussian density distribution (leftmost two figures). Shown are snapshots of the density field and probability currents with periodic boundary conditions at four representative times $t=0$, $t=\tau_B$, $t=5\,\tau_B$, and $t=50\,\tau_B$, where $\tau_B=\sigma^2/D_0$ denotes the characteristic diffusion time. 
			The top row corresponds to the normal (even) case $\kappa=0$, while the bottom row shows the odd-diffusive case with $\kappa=4$.
			The probability current is represented by arrows, whose direction indicates the local current orientation and whose color encodes the current magnitude $|\mathbf{J}|\sigma\tau_B$ (see color bar), whereas the local density $\rho\sigma^2$ is encoded in the background color using the inverse color scale (compare color bar in Fig.~\ref{figure1}).
			In the normal case, relaxation proceeds via purely radial currents flowing along density gradients. In contrast, odd diffusion generates pronounced transverse currents and circulating flow patterns during relaxation. Despite these qualitative differences in the transient dynamics, both systems relax toward the same homogeneous steady state at long times.}	
		\label{figure2}
	\end{figure*}
	Here, we formulate a generalized dynamical density functional theory for interacting odd-diffusive systems in the presence of external fields. Starting from the Smoluchowski description\cite{te2020classical}, we derive a closed evolution equation for the one-body density that consistently incorporates odd diffusion alongside conservative external potentials and pairwise interactions. This construction yields a minimal yet fully self-consistent odd-DDFT framework in which equilibrium structure is preserved by design, while the dynamics acquire additional transverse transport channels.

	To disentangle intrinsic odd-diffusive effects from geometric or boundary-induced influences, we first specialize this theory to a homogeneous bulk fluid by considering ultrasoft particles interacting via a Gaussian pair potential in the absence of confinement. In this setting, the theory provides a transparent description of how odd diffusion reshapes collective relaxation and probability currents in interacting fluids, while the steady-state density remains spatially uniform and indistinguishable from its equilibrium counterpart.
	
	Having established the bulk behavior within odd DDFT, we then introduce confinement to probe how spatial inhomogeneity and geometry interact with odd-diffusive dynamics. We focus on an azimuthally symmetric ring trap (see Fig.~\ref{figure1}), which generates a nonuniform 
	equilibrium density profile localized on a closed manifold. 
	Starting from a localized within this geometry, odd diffusion gives rise to pronounced transient probability currents that circulate along the ring during relaxation, allowing radial equilibration toward the trap minimum to be cleanly separated from tangential redistribution along the confinement. This setting provides a natural and unambiguous way to characterize angular transport and to quantify how odd diffusion reorganizes relaxation pathways.

	\section{Model and results}
	\begin{figure*}
		\centering
		\includegraphics[width=\linewidth]{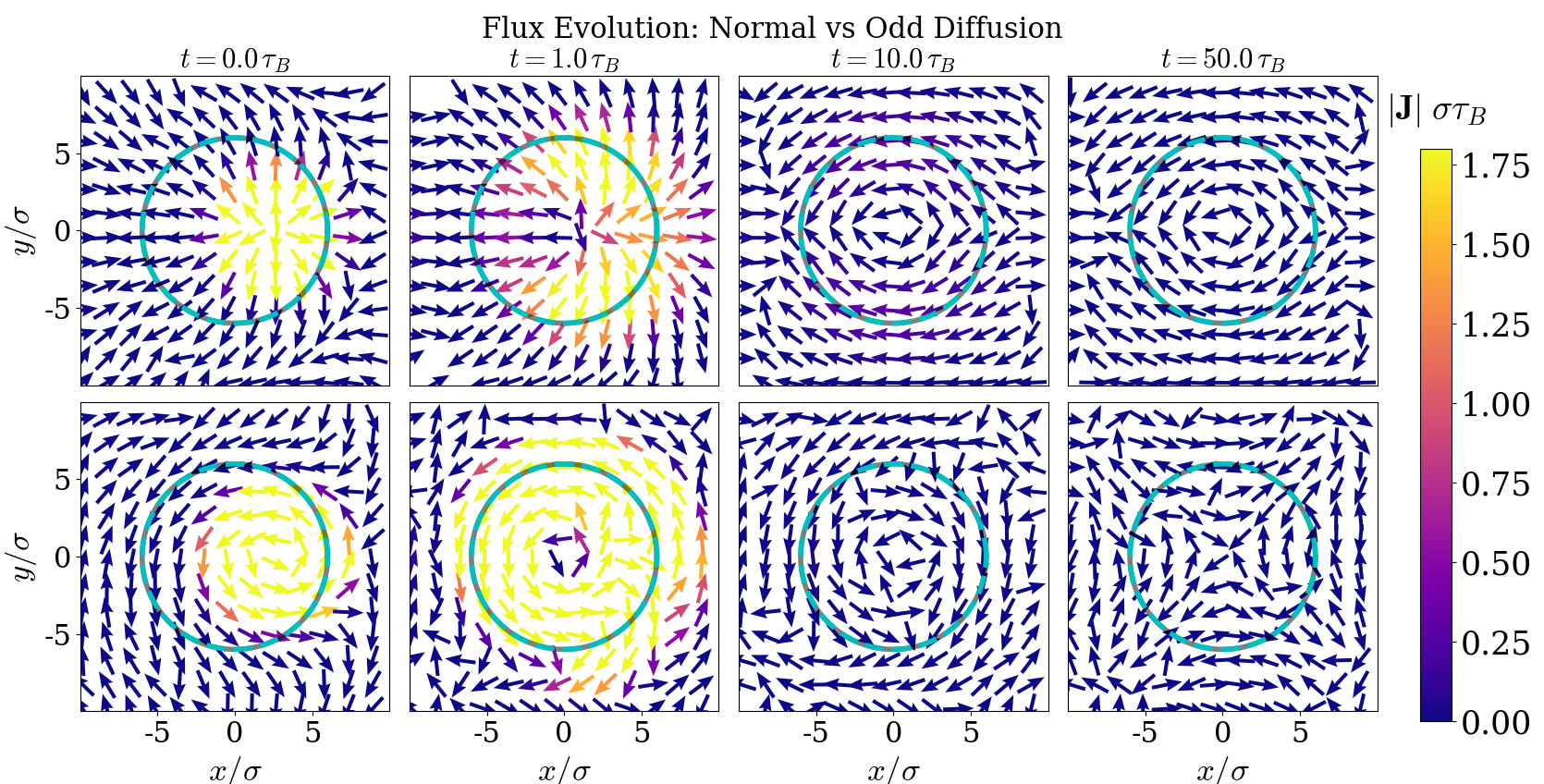}
		\caption{\textbf{Probability-current evolution in a harmonic ring trap for normal and odd diffusion.}
		 Time evolution of the one-body current field $\mathbf{J}(\mathbf{r},t)$ obtained from the
			odd-DDFT equation in a harmonic ring potential $V_{\rm ext}(r)=\tfrac12 k (r-R_0)^2$, where the particles are initially distributed in an off-center Gaussian blob.
			 Arrows indicate the local direction of the probability current, while their color encodes the current magnitude $|\mathbf{J}|\sigma\tau_B$ (see color bar).
			 The cyan dashed circle marks the ring radius $R_0=6.0\sigma$. 
			Columns correspond to $t=0$, $t=\tau_B$, $t=10\,\tau_B$, and $t=50\,\tau_B$
			where $\tau_B=\sigma^2/D_0$ denotes the characteristic diffusion time.
			 The top row corresponds to the normal case $\kappa=0$, for which relaxation is dominated by radial currents directed toward the ring minimum. The bottom row shows the odd-diffusive case $\kappa=4$, where pronounced circulating currents develop along the ring during intermediate stages of relaxation. Despite these transient angular fluxes, the probability current vanishes in the steady state, consistent with relaxation toward an equilibrium density profile.}
		\label{figure4}
	\end{figure*}
	We consider a two-dimensional system of $N$ interacting overdamped Brownian particles whose transport is governed by odd diffusion. Microscopically, odd diffusion is characterized by a diffusion tensor with an antisymmetric component, such that density gradients generate probability currents with a transverse contribution in addition to the usual down-gradient flux. The particles interact via a pairwise potential and may be subject to an external field, allowing us to describe both bulk and confined situations within a unified framework.
	
	For an isotropic two-dimensional system, the diffusion tensor takes the general form
	\begin{equation}
		\mathbf{D} = D_0\bigl(\mathbf{I} + \kappa \boldsymbol{\epsilon}\bigr),
		\label{eq:odd_tensor}
	\end{equation}
	where $D_0$ is the bare diffusion coefficient, $\mathbf{I}$ is the $2\times2$ identity matrix, and
	$\boldsymbol{\epsilon}=\bigl(\begin{smallmatrix}0 & 1\\ -1 & 0\end{smallmatrix}\bigr)$
	is the two-dimensional Levi--Civita tensor. The dimensionless parameter $\kappa$ controls the strength of the odd contribution. For $\kappa=0$, transport reduces to normal diffusion, whereas $\kappa\neq0$ induces transient probability currents perpendicular to density gradients without altering equilibrium states.
	
	\subsection{Many-body dynamics}
	\begin{figure*}[t]
		\centering
		\includegraphics[width=\textwidth]{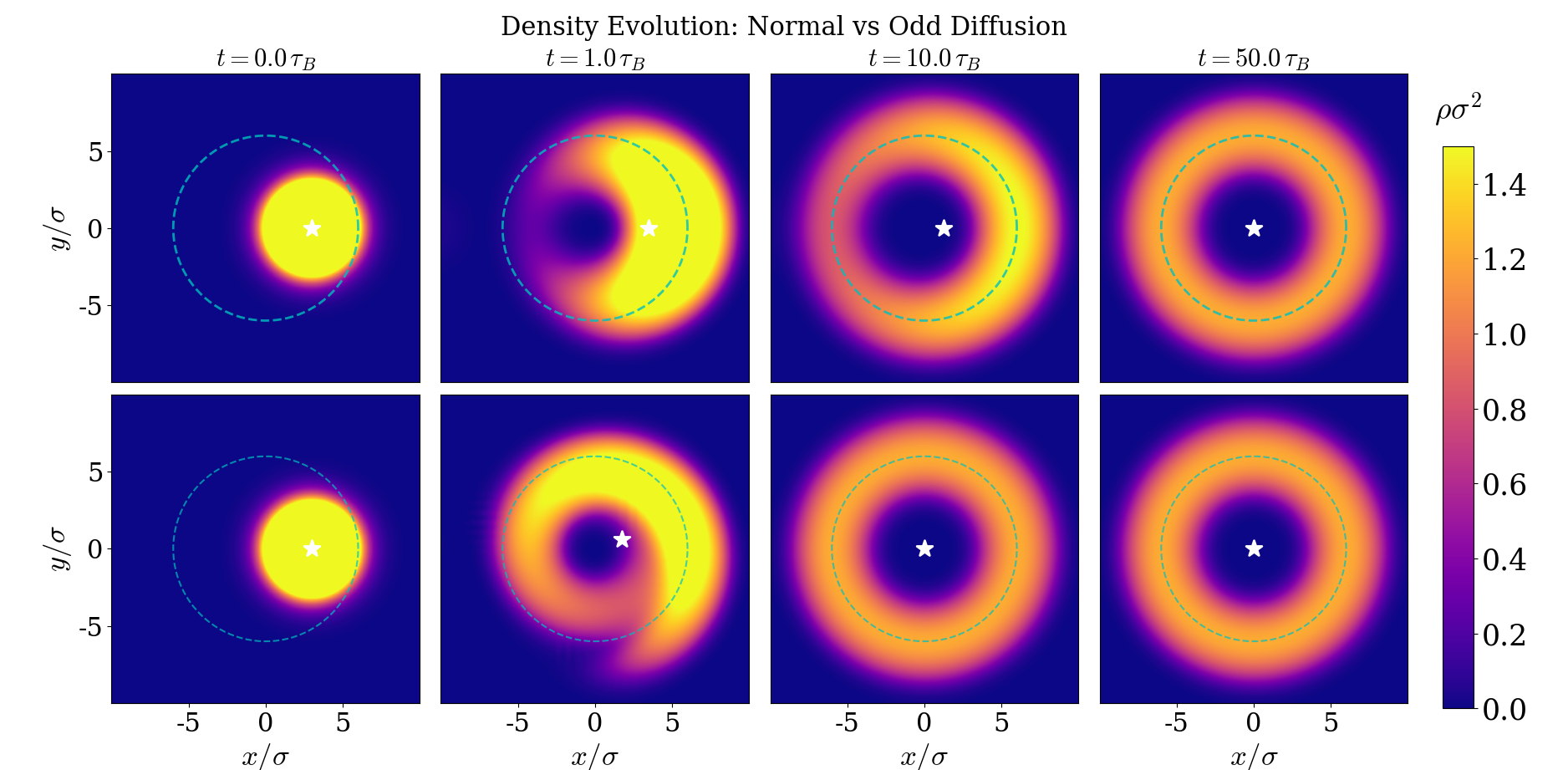}
		\caption{\textbf{Density evolution in a ring trap: normal vs odd diffusion.}
			Time evolution of the one-body density field $\rho(\mathbf{r},t)$ 
		 corresponding to the same system as in Fig.~\ref{figure3} at the same times.
			The cyan dashed circle marks the ring radius $R_0=6.0\sigma$.
			The star indicates the instantaneous density center-of-mass position
			$\mathbf{R}_{\rm CM}(t)$ (computed from the discretized density field).
			Odd diffusion generates a transverse, Hall-like component of the current, compare Fig.~\ref{figure4},
			leading to a pronounced angular advection of the density along the ring at intermediate
			times, while both cases ultimately relax towards a ring-shaped steady distribution  (see Appendix $C$ for comparison with simulation results).}
		\label{figure3}
	\end{figure*}
	The configurational probability density
	$P(\mathbf{R}^N,t)$ of particle positions
	$\mathbf{R}^N=(\mathbf{r}_1,\ldots,\mathbf{r}_N)$
	obeys the $N$-particle Smoluchowski equation
	\begin{equation}
		\partial_t P(\mathbf{R}^N,t)
		= -\sum_{i=1}^N \nabla_i \cdot \mathbf{J}_i(\mathbf{R}^N,t),
		\label{eq:many_body_fpe}
	\end{equation}
	where the probability current of particle $i$ is given by
	\begin{equation}
		\mathbf{J}_i
		 = -\mathbf{D}\cdot\left(\nabla_i P-\frac{1}{k_{\mathrm B}T}\cdot \mathbf{F}_i(\mathbf{R}^N)\,P\right).
	\end{equation}
	Here $\mathbf{F}_i=-\nabla_i U(\mathbf{R}^N)$ denotes the deterministic force acting on particle $i$, derived from the total potential energy
	\begin{equation}
		U(\mathbf{R}^N)
		=\sum_{i=1}^N V_{\mathrm{ext}}(\mathbf{r}_i)
		+\frac{1}{2}\sum_{i\neq j} V(|\mathbf{r}_i-\mathbf{r}_j|).
		\label{eq:tot_pot_en}
	\end{equation}
	The diffusion tensor $\mathbf{D}$ is taken to be constant in space and time and identical for all particles.
	
	\subsection{Odd dynamical density functional theory}

	Integrating Eq.~\eqref{eq:many_body_fpe} over $N-1$ particle coordinates yields a continuity equation 
		\begin{equation} \label{eq:cont_eq}
		\partial_t \rho(\mathbf{r},t)
		= -\nabla\cdot\mathbf{J}(\mathbf{r},t)
	\end{equation}
	for the one-body density field
	\begin{equation}
		\rho(\mathbf{r},t)
		= N\!\int\! \mathrm{d}\mathbf{r}_2\ldots\mathrm{d}\mathbf{r}_N\,
		P(\mathbf{r},\mathbf{r}_2,\ldots,\mathbf{r}_N;t).
	\end{equation}
However, the one-body current $\mathbf{J}$ still depends on the two-body density, which means that Eq.~\eqref{eq:cont_eq} is not closed. 
Dynamical density functional theory (DDFT) invokes the adiabatic approximation to close this hierarchy:  
at each time $t$, the nonequilibrium system is assumed to be described by an equilibrium free-energy functional $F[\rho]$ evaluated at the instantaneous density profile $\rho(\mathbf{r},t)$. This is achieved by introducing a local chemical potential
	\begin{equation}
		\mu(\mathbf{r},t)
		= \frac{\delta F[\rho]}{\delta\rho(\mathbf{r},t)},
	\end{equation}
 and applying it adiabatically in time, assuming that this generalization of the exact equilibrium condition still holds in a good approximation. Then,
	the one-body current can be expressed in closed form as
	\begin{equation}
		\mathbf{J}(\mathbf{r},t)
		= -\frac{\mathbf{D}}{k_{\mathrm B}T}\cdot
		\rho(\mathbf{r},t)\,\nabla\mu(\mathbf{r},t).
	\end{equation}
	The resulting odd-DDFT equation governing the density evolution reads (see Appendix $A$ for further details) 
	\begin{equation}
		\partial_t \rho(\mathbf{r},t)
		= \nabla\cdot
		\left[
		\frac{\mathbf{D}}{k_{\mathrm B}T}\cdot
		\rho(\mathbf{r},t)\,\nabla\mu(\mathbf{r},t)
		\right].
		\label{eq:odd_ddft}
	\end{equation}
	This equation has the same structure as standard DDFT, but with a diffusion tensor containing an antisymmetric (odd) contribution.
	
	\subsection{ Mean-field approximation for odd DDFT} 
	
 Within equilibrium density functional theory \cite{evans1979nature, tschopp2025routes, tschopp2025combining}, the Helmholtz free-energy functional is typically decomposed as
	\begin{equation}
		F[\rho]
		= F_{\mathrm{id}}[\rho]
		+ F_{\mathrm{ex}}[\rho]
		+ \int\! \mathrm{d}\mathbf{r}\,
		V_{\mathrm{ext}}(\mathbf{r})\,\rho(\mathbf{r}),
		\label{eq:free_energy}
	\end{equation}
	where the ideal-gas contribution  
	is given exactly by
	\begin{equation}
		F_{\mathrm{id}}[\rho]
		= k_{\mathrm B}T
		\int \mathrm{d}\mathbf{r}\,
		\rho(\mathbf{r})
		\left[
		\ln\!\left(\rho(\mathbf{r})\Lambda^2\right)-1
		\right],
	\end{equation}
	 with the thermal de Broglie wavelength $\Lambda$.
	 The excess free energy $F_{\mathrm{ex}}[\rho]$ describes the interactions between the particles and $V_{\mathrm{ext}}(\mathbf{r})$ denotes the one-body external potential, as introduced in Eq.~\eqref{eq:tot_pot_en}.

	For the interacting system considered here, we employ a mean-field (random-phase) approximation and use~\cite{dzubiella2003mean},
	\begin{equation}
		F_{\mathrm{ex}}[\rho]
		= \frac{1}{2}
		\int\!\!\int \mathrm{d}\mathbf{r}\,\mathrm{d}\mathbf{r}'\,
		V(|\mathbf{r}-\mathbf{r}'|)
		\rho(\mathbf{r})\rho(\mathbf{r}'),
		\label{eq:mf_excess}
	\end{equation}
	which is quasi-exact for ultrasoft interactions.
	Throughout this work, the pair potential is taken to be Gaussian~\cite{lang2000fluid, dzubiella2003mean},
	\begin{equation}
		V(r)
		= \varepsilon \exp\!\left[-\left(\frac{r}{\sigma}\right)^2\right],
	\end{equation}
	with $r=|\mathbf{r}|$, interaction strength $\varepsilon$ and range $\sigma$ (see inset of Fig.~\ref{figure1}).

 Inserting this free energy into the general odd-DDFT equation~\eqref{eq:odd_ddft}, we obtain the closed evolution equation 
		\begin{equation}
			\frac{\partial \rho(\mathbf{r},t)}{\partial t}
			= \nabla \cdot \left\{
			\mathbf{D}\cdot
			\left[
			\nabla \rho(\mathbf{r},t)
			+ \frac{\rho(\mathbf{r},t)}{k_{\mathrm B}T}
			\nabla \Phi(\mathbf{r},t)
			\right]
			\right\},
			\label{eq:complete_odd_ddft}
		\end{equation}
		where the total potential
		\begin{equation}
			\Phi(\mathbf{r},t)
			= \int \mathrm{d}\mathbf{r}'\,
			V(|\mathbf{r}-\mathbf{r}'|)
			\rho(\mathbf{r}',t)
			+ V_{\mathrm{ext}}(\mathbf{r}),
			\label{eq:total_potential}
		\end{equation}
		contains both the mean-field interaction contribution and the external confinement.

	\subsection{ Interacting odd-diffusive fluids in bulk}

	We first apply the odd-DDFT equation \eqref{eq:complete_odd_ddft} to an 
	interacting fluid in the absence of an external potential  $V_{\mathrm{ext}}(\mathbf{r})=0$,
	 to investigate how odd diffusion affects 
	the transient relaxation dynamics and probability currents  during the decay of a localized density profile to the homogeneous equilibrium state.
	All quantities appearing in this work, are expressed in dimensionless units with
	lengths measured in units of $\sigma$, energies in units of
	$k_{\mathrm B}T$, and time in units of the Brownian time
	$\tau_B=\sigma^2/D_0$.
	
In Fig.~\ref{figure2}, we compare the spatiotemporal evolution of both the density and probability currents for a nonconfined interacting fluid in the normal ($\kappa=0$) and odd-diffusive ($\kappa=4$) cases. While the density relaxes toward a homogeneous state in both systems, odd diffusion qualitatively reshapes the transient current patterns, giving rise to circulating probability fluxes during relaxation that are absent in the normal fluid.
In the Supplemental Information~\cite{SI}, we show videos of the density evolution and the corresponding probability fluxes in the system depicting how the nonconfined system relaxes towards the equilibrium (see Movie01 and Movie02 for normal diffusion and Movie03 and Movie04 for odd diffusion).
Taking a closer look at the density, we also notice that the relaxation is slightly accelerated in the odd case.

		\begin{figure}
			\centering
			\includegraphics[width=\linewidth]{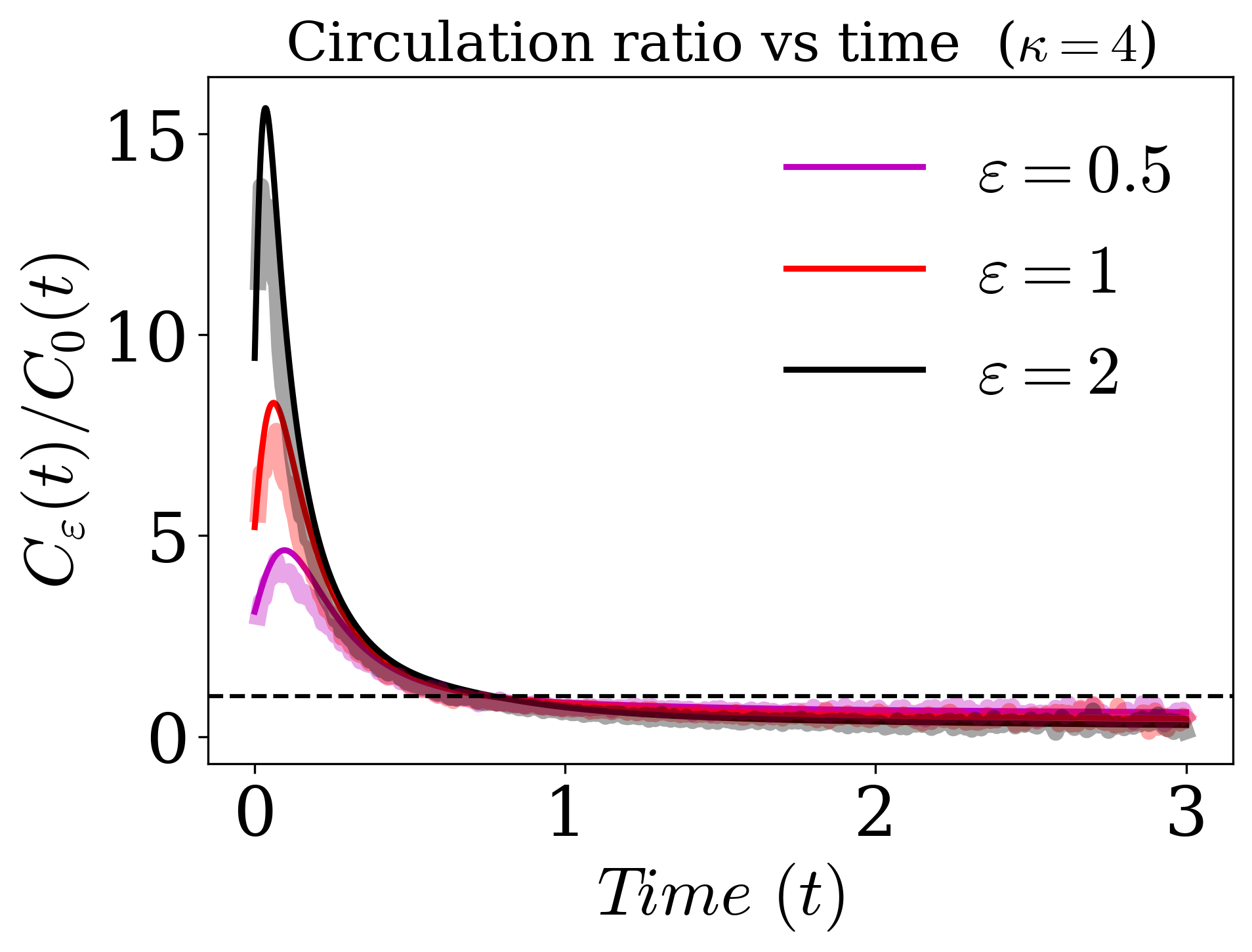}
			\caption{Interaction dependence of the normalized angular circulation in a confined odd-diffusive fluid.
				This figure shows the time evolution of the circulation ratio $C_\varepsilon(t)/C_0(t)$ for the odd-diffusive system ($\kappa=4$) at different interaction strengths $\varepsilon$, where $C_\varepsilon(t)$ denotes the angular circulation in the interacting system, given in Eq.~\eqref{eq:circulation}, and $C_0(t)$ its noninteracting counterpart (odd ideal gas). Normalizing by $C_0(t)$ isolates the effect of interactions on angular transport. Increasing interaction strength leads to a pronounced enhancement of the transient circulation peak, reflecting stronger collective redistribution along the ring during relaxation. At long times, the normalized circulation decays to zero for all $\varepsilon$, confirming that interactions modify only the transient dynamics while the equilibrium state remains free of persistent angular currents. Solid lines correspond to odd-DDFT predictions, while shaded bands represent simulation results.}

			\label{figure6}
		\end{figure}

				\begin{figure*}
	\centering
	\includegraphics[width=\linewidth]{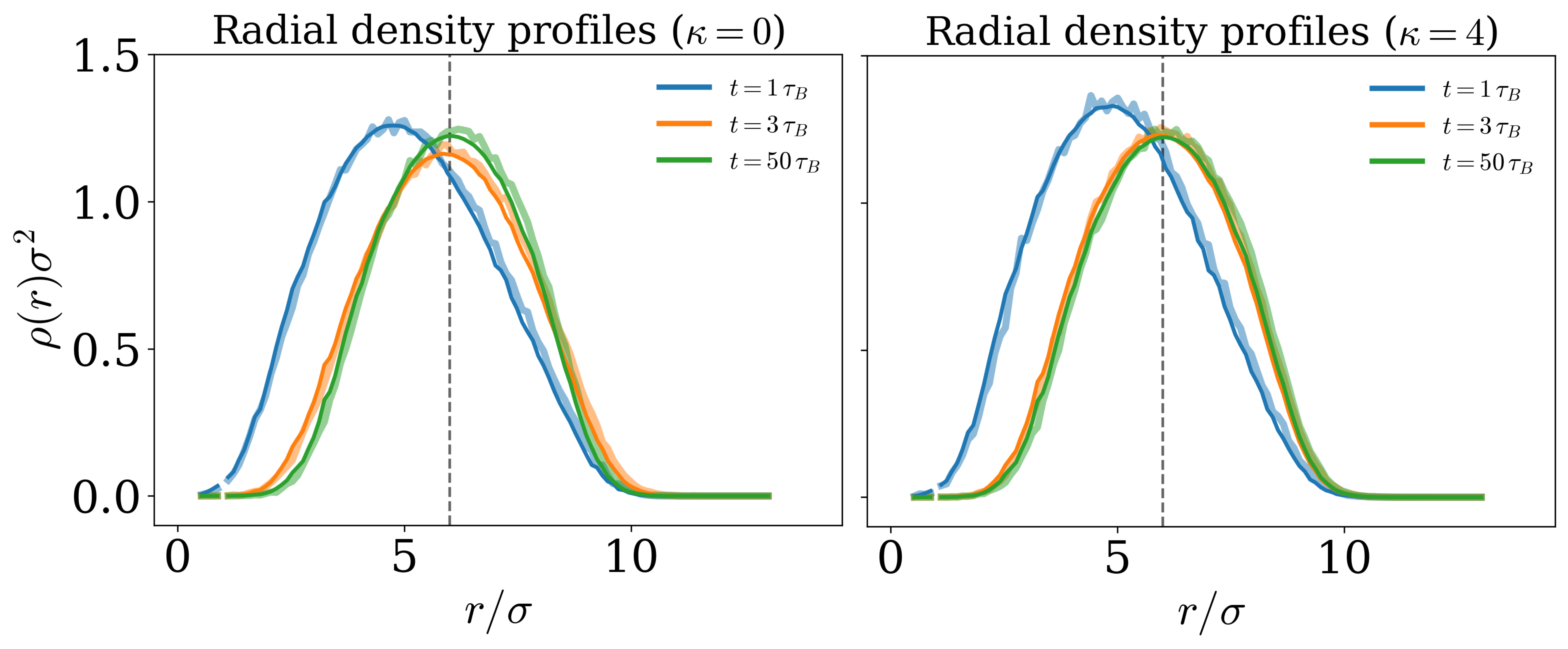}
	\caption{\textbf{Comparison between odd-DDFT theory and Brownian-dynamics simulations in a harmonic ring trap.} 
    Radial density profiles $\rho(r)$, obtained by azimuthal (angular) averaging of the instantaneous two-dimensional density field, for the normal case ($\kappa=0$, left) and the odd-diffusive case ($\kappa=4$, right) at representative times. Solid lines correspond to odd-DDFT predictions, while shaded bands represent simulation results (see Appendix $B$ and $C$ for details), reflecting statistical fluctuations. In both cases, the density relaxes toward the same ring-shaped equilibrium profile, demonstrating that odd diffusion does not affect static density distributions. The black, dashed line indicates the ring radius $R_0=6.0\sigma$.}
\label{figure5}
\end{figure*}

		\subsection{Confined odd-diffusive fluids}
		\label{subsec:confined}

		In the absence of external fields, odd diffusion modifies collective relaxation dynamics by inducing rotating probability currents. 
		To expose and quantify these chiral effects in more detail, 
		it is advantageous to consider a geometry that provides a natural notion of angular transport.
		We therefore introduce a radially symmetric ring confinement 
	   through the external potential
		\begin{equation}
			V_{\mathrm{ext}}(\mathbf{r})
			= \frac{1}{2}k\,(r-R_0)^2,
			\label{eq:ring_trap}
		\end{equation}
		where $r=|\mathbf{r}|=\sqrt{x^2+y^2}$ denotes the radial distance from the trap center, $R_0$ is the preferred ring radius, and $k$ controls the trap stiffness  (see Fig.~\ref{figure1}). This potential generates a spatially inhomogeneous equilibrium density, where the particles are confined to a narrow annular region around $R_0$, while preserving rotational symmetry (independent of the odd-diffusion parameter).

		 Our choice of this geometry serves two complementary purposes. First, it allows for an unambiguous definition of angular circulation associated with tangential probability currents. Second, it cleanly separates radial equilibration toward the preferred radius from tangential redistribution along the ring, thereby isolating the dynamical consequences of odd diffusion from purely relaxational effects. As we show below, this setting reveals how odd diffusion reshapes relaxation pathways  in a way such that the final equilibrium state is approached more rapidly.

In Fig.~\ref{figure4}, we show the time evolution of the one-body current in a harmonic ring trap for normal and odd diffusion. Similar to the bulk case, odd diffusion gives rise to pronounced transient probability currents circulating along the ring during relaxation, whereas the normal system exhibits purely relaxational currents directed toward the trap minimum. Remarkably, the circulating currents in the odd-diffusive case  also lead to an angular redistribution of the one-body density during relaxation, as illustrated in Fig.~\ref{figure3}. 
Although the system ultimately relaxes toward the same ring-shaped equilibrium density profile as for normal diffusion, the altered pathway enables a notably faster equilibration, which becomes apparent from position of the center of mass (compare the stars in Fig.~\ref{figure3}).
		In the Supplemental Information~\cite{SI} we show videos of the density evolution and the corresponding probability fluxes, 
		depicting how the confined system relaxes towards the equilibrium with and without odd diffusion (see Movie05 and Movie06 for normal diffusion and Movie07 and Movie08 for odd diffusion).

		To characterize the observed behavior quantitatively  and investigate the role of the interactions, we introduce a measure for angular circulation 
		that captures the net tangential probability current integrated along the ring. The circulation is defined as the line integral of the tangential component of the probability current along the ring:
		\begin{equation}
		C_\varepsilon(t)=\oint_{r=R_0} \mathbf{J}(\mathbf{r},t)\cdot \hat{\boldsymbol{\theta}}\,\mathrm{d}l,
		\label{eq:circulation}
		\end{equation}
		where $\hat{\boldsymbol{\theta}}$ denotes the local azimuthal unit vector and $\mathrm{d}l$ is the line element along the ring. 
		As shown in Fig.~\ref{figure6}, 
		this circulation exhibits a transient peak which 
		decays to zero at long times. The vanishing of circulation in the steady state confirms that the confined odd-diffusive fluid relaxes to thermal equilibrium and does not sustain persistent edge currents, in accordance with equilibrium statistical mechanics.
		To highlight the role of interactions,  we have normalized the circulation by its value for an odd ideal gas in Fig.~\ref{figure6}. 
		We observe that the magnitude of the transient circulation strongly increases with increasing repulsion strength, while the normalized measure collapses at long times. 
		 This implies that interactions further enhance the accelerated relaxation due to odd diffusivity.


To demonstrate the quantitative accuracy of our odd DDFT, we compare its predictions with Brownian-dynamics simulations  and find an excellent agreement for the angular circulation in Fig.~\ref{figure6}.
	We further show in Fig.~\ref{figure5} that the radial density profiles show excellent agreement at all times for both confined normal and odd-diffusive fluids.

		\section{Discussion}
		
		In this work, we have formulated a dynamical density functional theory for interacting odd-diffusive fluids and used it to clarify how transverse diffusion reshapes collective relaxation dynamics in bulk and under confinement. A central outcome is the robust separation between static and dynamical properties: odd diffusion leaves equilibrium density distributions unchanged, while inducing pronounced transient probability currents and chiral relaxation pathways. This separation allows nonequilibrium transport phenomena to be isolated and analyzed without modifying the underlying equilibrium structure.

		For confined systems, the harmonic ring geometry provides a minimal and transparent setting to reveal these effects. Odd diffusion generates circulating probability currents during relaxation and leads to angular redistribution of density along the ring, yet these currents decay at long times and do not result in persistent steady-state transport. Repulsive interactions enhance the magnitude 
		of the transient circulation, promoting a faster relaxation to equilibrium, which is, in turn, accompanied by the gradual disappearance of the circulation. 
		 The quantitative agreement between odd-DDFT predictions and Brownian-dynamics simulations demonstrates that both the mean-field treatment of interactions and the adiabatic approximation remain reliable in the present system, even when relaxation dynamics become strongly chiral.
		
		The fact that an adiabatic DDFT is capable of predicting a genuine nonequilibrium phenomenon like odd circulation with quantitative accuracy is, at first glance, quite surprising.
		It is well known, that neglecting the so-called superadiabatic forces that emerge in generic interacting systems out of equilibrium \cite{PhysRevLett.113.167801} can lead to a quite poor performance of DDFT.
		One prominent example is a fluid under shear, where standard DDFT fails even on a qualitative level unless being supplemented by an empirical flow kernel or explicit superadiabatic terms \cite{brader2011density, scacchi2016driven, scacchi2018flow, tschopp2024superadiabatic}.
		A key difference to the system at hand is that shear enters through an additional (nonconservative) external force term, whereas odd-diffusivity 
		is represented by a modified diffusion tensor, thereby merely transforming rather than coupling to the (adiabatic) interaction forces.
		 It would be interesting to include odd diffusivity in the investigation of density--potential mappings \cite{klatt2023foundation} or to incorporate it
		 into the frameworks of superadiabtic DDFT \cite{tschopp2022first}  or power functional theory \cite{schmidt2013power, schmidt2022power}  to better characterize the (supposedly minor) role of odd superadiabatic forces in more detail.

		Several natural extensions of the present framework emerge. First, odd diffusion is expected to influence ordering and crystallization kinetics in dense systems, even though equilibrium structures remain unchanged. Extending odd DDFT to crystalline or near-crystalline regimes would allow one to study how transverse diffusion modifies defect dynamics, nucleation pathways, and stress relaxation during ordering~\cite{van2008colloidal, van2013vacancy}. Second, incorporating hydrodynamic interactions represents an important direction, as long-ranged hydrodynamic couplings are known to affect collective relaxation and transport in confined fluids; their interplay with odd diffusion may give rise to new classes of transient flow patterns and correlations~\cite{rex2009dynamical}. Finally, odd diffusion is closely connected to systems driven by multiple or anisotropic noise sources~\cite{filliger2007brownian, abdoli2022tunable, abdoli2022escape, abdoli2025enhanced}. Recent work has shown that orthogonal noise components can generate higher-order circulating modes and structured probability currents in overdamped systems~\cite{abdoli2025quadrupolar}. Developing a density-functional description for interacting fluids subjected to such noise structures would provide a unified framework for studying odd diffusion, anisotropic fluctuations, and their collective consequences.

		\begin{acknowledgements}
	We gratefully acknowledge financial support from the Deutsche Forschungsgemeinschaft (DFG) under Project No.~556762905 (AB~1083/1--1) for I.\ A.\ and
	  through the SPP 2265 under Grant Nos.\ WI 5527/1-2 for R.\ W.\ and LO 418/25-2 for H.\ L. 

		\end{acknowledgements}
		
		\section*{AUTHOR DECLARATIONS}
		\subsection*{Conflict of Interest}
		The authors have no conflicts to disclose.
		
		\section*{DATA AVAILABILITY}
		The data that support the findings of this study are available
		from the corresponding author upon reasonable request.
		\appendix
		
		\section*{Appendices}
		\section*{Appendix A: Derivation of the odd-DDFT equation}
		
		In this Appendix, we derive the dynamical density functional theory (DDFT) equation employed throughout the main text, starting from the $N$-particle Smoluchowski equation and applying the standard adiabatic DDFT closure. The derivation is carried out for a general diffusion tensor containing an antisymmetric (odd) contribution, thereby establishing the odd-DDFT framework used in our analysis.
		
		
		We consider $N$ overdamped Brownian particles in two spatial dimensions with positions
		$\mathbf{R}^N=(\mathbf{r}_1,\dots,\mathbf{r}_N)$. The total potential energy of the system is
		\begin{equation}
			U_N(\mathbf{R}^N)
			= \sum_{i=1}^N V_{\mathrm{ext}}(\mathbf{r}_i)
			+ \frac{1}{2}\sum_{i\neq j} V(|\mathbf{r}_i-\mathbf{r}_j|),
			\label{eq:App_UN}
		\end{equation}
		where $V_{\mathrm{ext}}$ denotes the external potential and $V(r)$ the pair interaction.
		
		The configurational probability density $P(\mathbf{R}^N,t)$ obeys the Smoluchowski equation
		\begin{equation}
			\partial_t P(\mathbf{R}^N,t)
			= - \sum_{i=1}^N \boldsymbol{\nabla}_i \cdot \mathbf{J}_i(\mathbf{R}^N,t),
			\label{eq:App_Smol}
		\end{equation}
		with single-particle probability currents
		\begin{equation}
			\mathbf{J}_i
			= - \mathbf{D} \cdot
			\left[
			\boldsymbol{\nabla}_i P
			+ \beta P\,\boldsymbol{\nabla}_i U_N
			\right],
			\label{eq:App_Ji}
		\end{equation}
		where $\beta=(k_{\mathrm B}T)^{-1}$. The diffusion tensor $\mathbf{D}$ is taken to be constant, identical for all particles, and decomposed into symmetric and antisymmetric parts,
		\begin{equation}
			\mathbf{D} = \mathbf{D}^{\mathrm S} + \mathbf{D}^{\mathrm A},
			\qquad
			\mathbf{D}^{\mathrm A}=-(\mathbf{D}^{\mathrm A})^{\mathsf T}.
			\label{eq:App_Ddecomp}
		\end{equation}
		The antisymmetric contribution encodes odd diffusion and breaks time-reversal symmetry at the level of transport.
		
		
		The one-body density field is defined as
		\begin{equation}
			\rho(\mathbf{r},t)
			= N \int \mathrm{d}\mathbf{r}_2\dots\mathrm{d}\mathbf{r}_N\,
			P(\mathbf{r},\mathbf{r}_2,\dots,\mathbf{r}_N;t).
			\label{eq:App_rho}
		\end{equation}
		Integrating Eq.~\eqref{eq:App_Smol} over $\mathbf{r}_2,\dots,\mathbf{r}_N$ yields the exact continuity equation
		\begin{equation}
			\partial_t \rho(\mathbf{r},t)
			= - \boldsymbol{\nabla} \cdot \mathbf{J}(\mathbf{r},t),
			\label{eq:App_cont}
		\end{equation}
		with the one-body probability current
		\begin{equation}
			\mathbf{J}(\mathbf{r},t)
			= N \int \mathrm{d}\mathbf{r}_2\dots\mathrm{d}\mathbf{r}_N\,
			\mathbf{J}_1(\mathbf{R}^N,t).
			\label{eq:App_J1}
		\end{equation}
		
		Substituting Eq.~\eqref{eq:App_Ji} and carrying out the integrations yields
		\begin{align}
			\mathbf{J}(\mathbf{r},t)
			&= - \mathbf{D} \cdot
			\Bigl[
			\boldsymbol{\nabla}\rho(\mathbf{r},t)
			+ \beta\,\rho(\mathbf{r},t)\boldsymbol{\nabla}V_{\mathrm{ext}}(\mathbf{r})
			\nonumber\\
			&\quad
			+ \beta \!\int\!\mathrm{d}\mathbf{r}'\,
			\rho^{(2)}(\mathbf{r},\mathbf{r}',t)
			\boldsymbol{\nabla}V(|\mathbf{r}-\mathbf{r}'|)
			\Bigr],
			\label{eq:App_Jrho2}
		\end{align}
		where $\rho^{(2)}(\mathbf{r},\mathbf{r}',t)$ denotes the two-body density. Equations
		\eqref{eq:App_cont}–\eqref{eq:App_Jrho2} are exact but not closed. DDFT closes the hierarchy by invoking the adiabatic approximation: at each time $t$, the nonequilibrium system is approximated by an equilibrium system with the same instantaneous density profile $\rho(\mathbf{r},t)$. Using the equilibrium sum rule of classical density functional theory and applying it adiabatically yields
		\begin{equation}
			\int \mathrm{d}\mathbf{r}'\,
			\rho^{(2)}(\mathbf{r},\mathbf{r}',t)
			\boldsymbol{\nabla}V(|\mathbf{r}-\mathbf{r}'|)
			\simeq
			\rho(\mathbf{r},t)\boldsymbol{\nabla}\mu_{\mathrm{ex}}(\mathbf{r},t).
			\label{eq:App_sumrule}
		\end{equation}
		
		Inserting Eq.~\eqref{eq:App_sumrule} into Eq.~\eqref{eq:App_Jrho2} and then into Eq.~\eqref{eq:App_cont} gives rise to the final odd-DDFT equation given in the main text.

\section*{Appendix B: Numerical solution of the odd-DDFT equation}

In this Appendix we describe the numerical scheme used to solve the
mean-field odd-DDFT equation for the one-body density field
$\rho(\mathbf r,t)$,
\begin{equation}
	\partial_t \rho(\mathbf r,t)
	= \boldsymbol{\nabla}\cdot
	\left[
	\mathbf D
	\left(
	\boldsymbol{\nabla}\rho(\mathbf r,t)
	+ \frac{\rho(\mathbf r,t)}{k_{\mathrm B}T}\,
	\boldsymbol{\nabla}\Phi(\mathbf r,t)
	\right)
	\right],
	\label{eq:B_ddft}
\end{equation}
where the odd diffusion tensor in two dimensions is
\begin{equation}
	\mathbf D
	= D_0\bigl(\mathbf 1+\kappa\,\boldsymbol{\epsilon}\bigr)
	=
	\begin{pmatrix}
		D_0 & \kappa D_0 \\
		-\kappa D_0 & D_0
	\end{pmatrix}.
	\label{eq:B_D}
\end{equation}
Here $D_0$ is the normal diffusion coefficient, $\kappa$ is the
dimensionless oddness parameter, and $\boldsymbol{\epsilon}$ is the
two-dimensional Levi--Civita tensor. The total potential
\begin{equation}
	\Phi(\mathbf r,t)
	= \Phi_{\mathrm{MF}}(\mathbf r,t)
	+ V_{\mathrm{ext}}(\mathbf r)
	\label{eq:B_Phi}
\end{equation}
contains both mean-field interaction and external contributions.

\subsection*{B.1 \quad Mean-field convolution}
	\begin{figure*}[t]
	\centering
	\includegraphics[width=\textwidth]{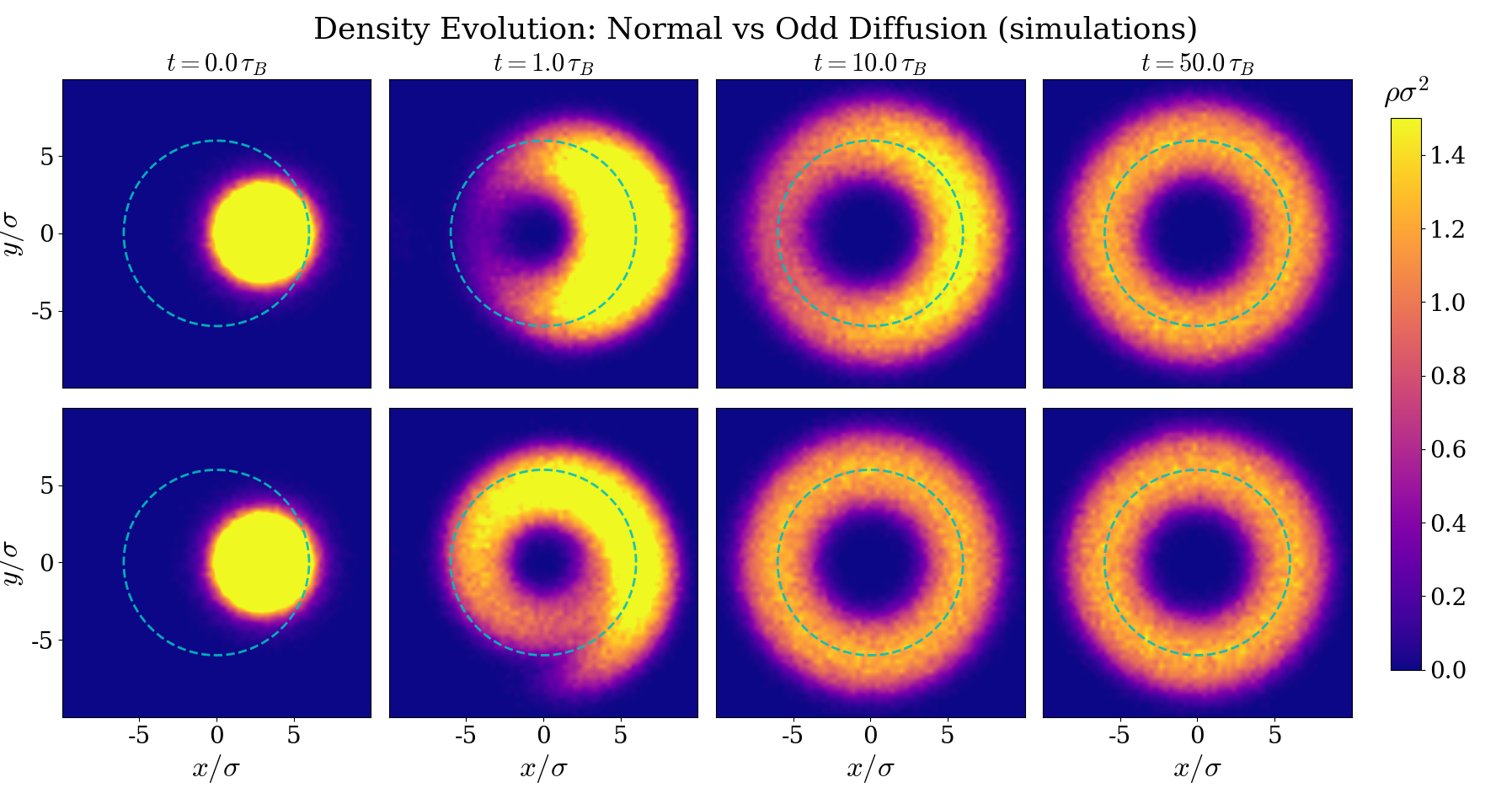}
	\caption{\textbf{Simulations: density evolution in a ring trap: normal vs odd diffusion.} 
		Time evolution of the one-body density field $\rho(\mathbf{r},t)$ obtained from the
		Brownian dynamics simulations for the same system as in Fig.~3 in the main text, in a harmonic ring potential $V_{\rm ext}(r)=\tfrac12 k (r-R_0)^2$.
		The top row shows the \emph{normal} (even) case $\kappa=0$, while the bottom row shows
		the \emph{odd} case $\kappa=4$.
		Columns correspond to $t=0$, $t=\tau_B$, $t=10\,\tau_B$, and $t=50\,\tau_B$
		where $\tau_B=\sigma^2/D_0$ denotes the characteristic diffusion time.
		The cyan dashed circle marks the ring radius $R_0=6.0\sigma$.
		Odd diffusion generates a transverse, Hall-like component of the current,
		leading to a pronounced angular advection of the density along the ring at intermediate
		times, while both cases ultimately relax towards a ring-shaped steady distribution. The simulation results are in an excellent agreement with those from the odd-DDFT.}
	\label{fig:density_evolution_app}
\end{figure*}

The mean-field interaction contribution is given by the convolution
\begin{equation}
	\Phi_{\mathrm{MF}}(\mathbf r,t)
	= \int \mathrm d\mathbf{r}'\,
	V(|\mathbf r-\mathbf r'|)\rho(\mathbf r',t),
\end{equation}
with a Gaussian pair potential
$V(r)=\varepsilon\exp[-(r/\sigma)^2]$.
This convolution is evaluated spectrally using fast Fourier transforms
(FFTs). To this end, the pair potential is discretized on the same
periodic grid using a minimum-image convention to construct a periodic
kernel $V_{\mathrm{per}}(\mathbf r)$. The kernel is shifted such that
$r=0$ corresponds to the $(0,0)$ grid index, and its discrete Fourier
transform $\widetilde V_{\mathrm{per}}(\mathbf k)$ is precomputed. The
mean-field potential is then obtained as
\begin{equation}
	\Phi_{\mathrm{MF}}(\mathbf r,t)
	= \mathcal F^{-1}\!\left[
	\widetilde V_{\mathrm{per}}(\mathbf k)\,
	\widetilde\rho(\mathbf k,t)
	\right],
\end{equation}
where $\mathcal F$ denotes the discrete Fourier transform on the
periodic domain.

Spatial gradients of $\mu$ are computed using centered finite
differences with periodic wrapping. The probability current
$\mathbf J=-\boldsymbol{\Gamma}\rho\nabla\mu$,
with $\boldsymbol{\Gamma}=\mathbf D/(k_{\mathrm B}T)$, is evaluated in
component form as
\begin{align}
	J^x_{i,j}
	&= -\frac{D_0}{k_{\mathrm B}T}
	\bigl[
	\rho_{i,j}(\partial_x\mu)_{i,j}
	- \kappa\,\rho_{i,j}(\partial_y\mu)_{i,j}
	\bigr],
	\\
	J^y_{i,j}
	&= -\frac{D_0}{k_{\mathrm B}T}
	\bigl[
	\kappa\,\rho_{i,j}(\partial_x\mu)_{i,j}
	+ \rho_{i,j}(\partial_y\mu)_{i,j}
	\bigr],
\end{align}
which explicitly separates the normal diffusive and transverse odd
contributions.

\subsection*{B.2 \quad Conservative update, time stepping, and Initial condition}

The continuity equation
$\partial_t\rho+\boldsymbol{\nabla}\cdot\mathbf J=0$
is discretized in finite-volume form to ensure exact conservation of
particle number. The discrete divergence at cell $(i,j)$ is evaluated
from face-centered fluxes obtained by arithmetic averaging of adjacent
cell-centered values. Time integration is performed using an explicit
Euler scheme,
\begin{equation}
	\rho^{n+1}_{i,j}
	= \rho^n_{i,j}
	- \Delta t\,
	(\boldsymbol{\nabla}\cdot\mathbf J^n)_{i,j}.
\end{equation}
After each update we enforce positivity by projecting
$\rho^{n+1}_{i,j}\leftarrow\max(\rho^{n+1}_{i,j},0)$.

The time step $\Delta t$ is chosen to satisfy a conservative stability
bound for the diffusive operator with odd corrections,
\begin{equation}
	\Delta t
	< \frac{\min(\Delta x^2,\Delta y^2)}
	{4D_0(1+\kappa^2)}.
\end{equation}


The initial condition is a Gaussian density profile displaced from the
center of the trap,
\begin{equation}
	\rho(\mathbf r,0)
	= \frac{N}{2\pi\sigma_r^2}
	\exp\!\left[-\frac{|\mathbf r-\mathbf r_0|^2}{2\sigma_r^2}\right],
\end{equation}
normalized such that
$\int\mathrm d^2 r\,\rho(\mathbf r,0)=N$.
During the simulation we monitor conservation of the total particle
number, the center-of-mass position, azimuthally averaged radial density
profiles, and the probability current field. The angular circulation is
computed from the tangential component of the current along the ring, as
defined in the main text.

\section*{Appendix C: Langevin dynamics simulations in the small-mass limit}

We validate the odd-DDFT predictions by Brownian dynamics simulations of \(N\) interacting
particles in a periodic square box of size \(L_x\times L_y\), subject to the same ring trap and
Gaussian-core interactions as in the DDFT. 

\subsection*{C.1 Equations of motion}

Each particle \(i\) obeys the underdamped Langevin dynamics
\begin{align}
	\dot{\mathbf{r}}_i &= \mathbf{v}_i,\\
	m\,\dot{\mathbf{v}}_i &=
	-\gamma\,(I-\kappa\boldsymbol{\varepsilon})\,\mathbf{v}_i
	+ \mathbf{F}_i(\{\mathbf{r}_j\})
	+ \sqrt{2\gamma}\,\boldsymbol{\eta}_i(t),
	\label{eq:BD_underdamped}
\end{align}
where \(\boldsymbol{\varepsilon}\) is the 2D antisymmetric tensor and
\(\langle \eta_{i\alpha}(t)\eta_{j\beta}(t')\rangle=\delta_{ij}\delta_{\alpha\beta}\delta(t-t')\).
The total force is \(\mathbf{F}_i=\mathbf{F}_i^{\rm int}+\mathbf{F}_i^{\rm ext}\), with
\begin{equation}
	V(r)=\varepsilon\,e^{-r^2},\qquad
	V_{\rm ext}(r)=\frac{k}{2}(r-R_0)^2.
\end{equation}

\subsection*{C.2 Small-mass regularization}

To access the overdamped regime in a controlled way for all \(\kappa\), we use the
\(\kappa\)-dependent rescaling
\begin{equation}
	\gamma(\kappa)=\frac{\gamma_0}{1+\kappa^2},
	\qquad
	m(\kappa)=\frac{m_0}{1+\kappa^2},
	\label{eq:small_mass_scaling}
\end{equation}
which keeps the inertial relaxation time \(\tau_m=m/\gamma=m_0/\gamma_0\) constant and
prevents an artificial change of the friction-matrix norm with increasing oddness~\cite{kalz2025reversal}.

\subsection*{C.3 Parameters and initialization}

The time step is \(\Delta t=10^{-5}\) and we use the system size $L_x=L_y=20$, and the number of particles $N=200$. For the small-mass regularization we set \(\gamma_0=1\) and \(m_0=5\times 10^{-3}\), so that
\(\tau_m=m/\gamma=5\times10^{-3}\ll 1\).
Initial positions are sampled from a Gaussian blob centered at
\((x_0,y_0)=(3,0)\) with width \(\sigma_r=1.5\), and initial velocities are drawn from the
Maxwell--Boltzmann distribution with variance \(k_BT/m\) per component.
Ensemble averages are obtained over \(n_{\rm real}\) independent realizations
(\(n_{\rm real}=5000\)), using independent random seeds.

		\section*{References}
%
	\end{document}